\documentclass[final,1p,times]{elsarticle} 
\usepackage{graphicx}
\usepackage{amssymb} 
\usepackage{amsthm} 
\usepackage{lineno}

\newcommand{\pT}{$p_{T}$ }
\newcommand{\Jpsi}{ J/$\psi$ }

\journal{Nuclear Physics A} 

\begin{document}

\begin{frontmatter} 

% Your Title - please insert
\title{Quarkonia production in the STAR experiment}

%% Single author (and collaboration) - please insert
\author{Barbara Trzeciak for the STAR\fnref{col1} Collaboration}
\fntext[col1] {A list of members of the STAR Collaboration and acknowledgements can be found at the end of this issue.}
\address{Faculty of Physics, Warsaw University of Technology \\ Koszykowa 75, 00-662 Warsaw, Poland}

\begin{abstract} 
In this proceedings the recent STAR results of \Jpsi and $\Upsilon$ production in $p+p$, $d$+Au and Au+Au collisions at $\sqrt{s_{NN}}$ = 200 GeV at mid-rapidity are reported. \Jpsi \pT spectra in $p+p$ and Au+Au collisions for both low and high \pT are shown. \Jpsi nuclear modification factor ($R_{AA}$) in $d$+Au and Au+Au collisions and $\Upsilon$ $R_{AA}$ in Au+Au collisions are reported. Also, \Jpsi polarization in $p+p$ collisions and \Jpsi $v_{2}$ for semi-central Au+Au collisions are presented.
   
\end{abstract} 

\end{frontmatter} % do not change

%% linenumbers are useful for reviewing process
%\linenumbers

\section{Introduction}

The suppression of quarkonia (charmonia and bottomonia) production in high energy nuclear collisions relative to $p+p$ collisions, due to the Debye screening of the quark-antiquark potential, was proposed as a signature of the formation of QGP \cite{MatsuiSatz}. However, there are other effects that may affect the observed quarkonia production. The cold nuclear matter effects, e.g. nuclear shadowing, Cronin effect, nuclear absorption, can be tested in $p$+A or $d$+A collisions. The other hot nuclear effects, such as recombination of quark-antiquark pairs might be also present. The interpretation of the quarkonia modification in QGP requires also understanding of the quarkonia production mechanism in $p+p$ collisions. At RHIC energies the $\Upsilon$ meson is a cleaner probe comparing to \Jpsi due to negligible contributions from $b$-$\bar{b}$ recombination and non-thermal suppression from co-mover absorption. Measurements of the quarkonia production in different colliding systems, centralities and collision energies are needed to understand those effects. In this proceedings results on \Jpsi  production in $p+p$, $d$+Au and Au+Au collisions and $\Upsilon$ production in $p+p$ and Au+Au collisions via the dielectron decay channel at mid-rapidity at $\sqrt{s_{NN}}$ = 200 GeV in the STAR experiment are presented.

\vspace{-10pt}
\section{$J/\psi$ production and polarization in $p+p$ collisions at 200 GeV}

\begin{figure}[htbp]
	\begin{minipage}[b]{0.49\linewidth}
	\vspace{-10pt}
		\centering
		\includegraphics[width=0.85\textwidth]{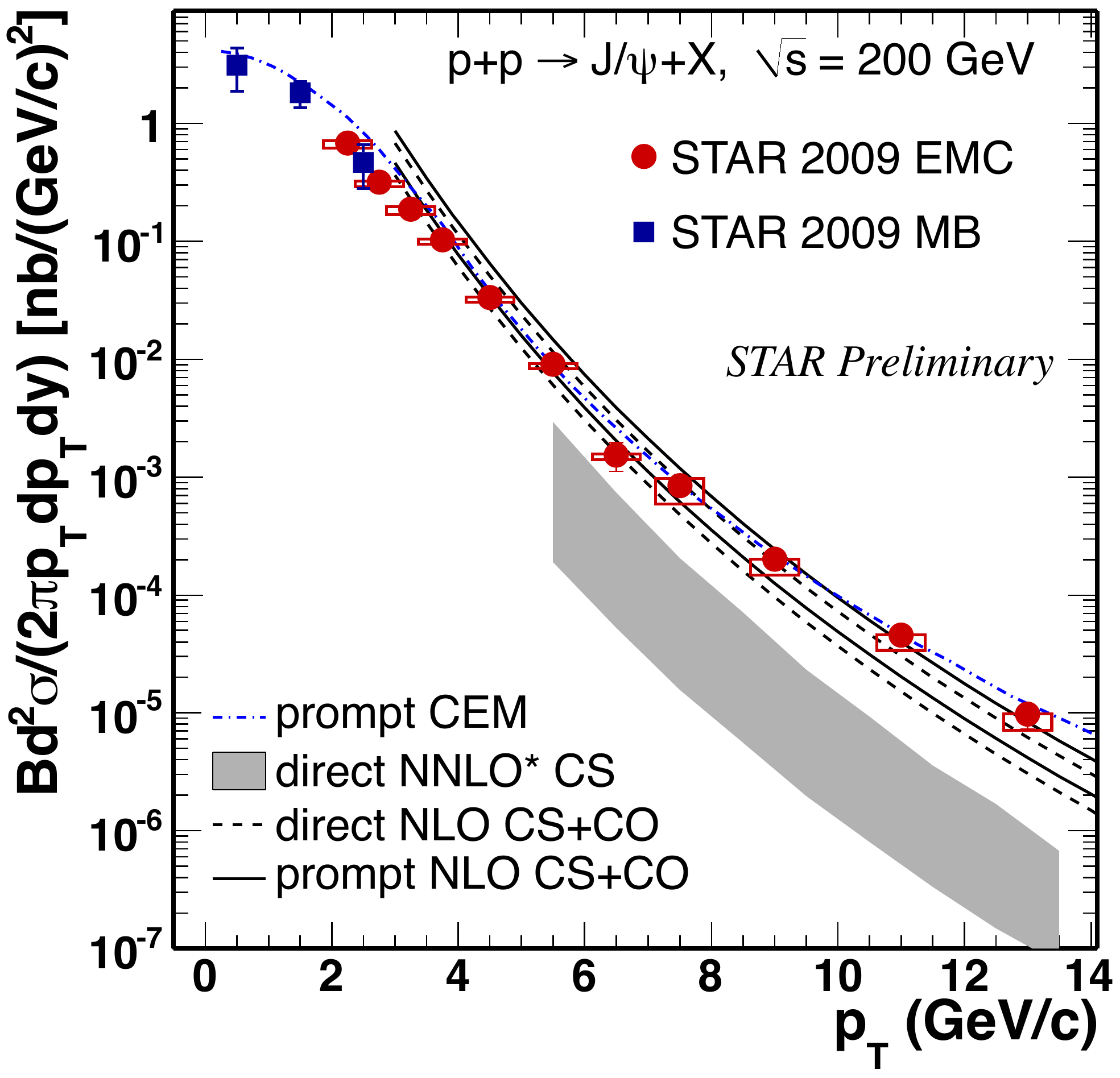}
		\vspace{-15pt}
		\caption{$J/\psi$ invariant cross section vs. \pT in $p+p$ collisions. Blue rectangles and red circles represent low-\pT  and  high-\pT \cite{STARhighPtJpsi} measurements, respectively. Upper and lower curves for solid and dashed lines represent upper and lower limits of a model prediction. }
		\vspace{-10pt}
		\label{fig:JpsiPtSpectrum_pp}
	\end{minipage}
	\hspace{0.cm}
	\begin{minipage}[b]{0.49\linewidth}
	\vspace{-10pt}
		\centering
		\includegraphics[width=1.\linewidth]{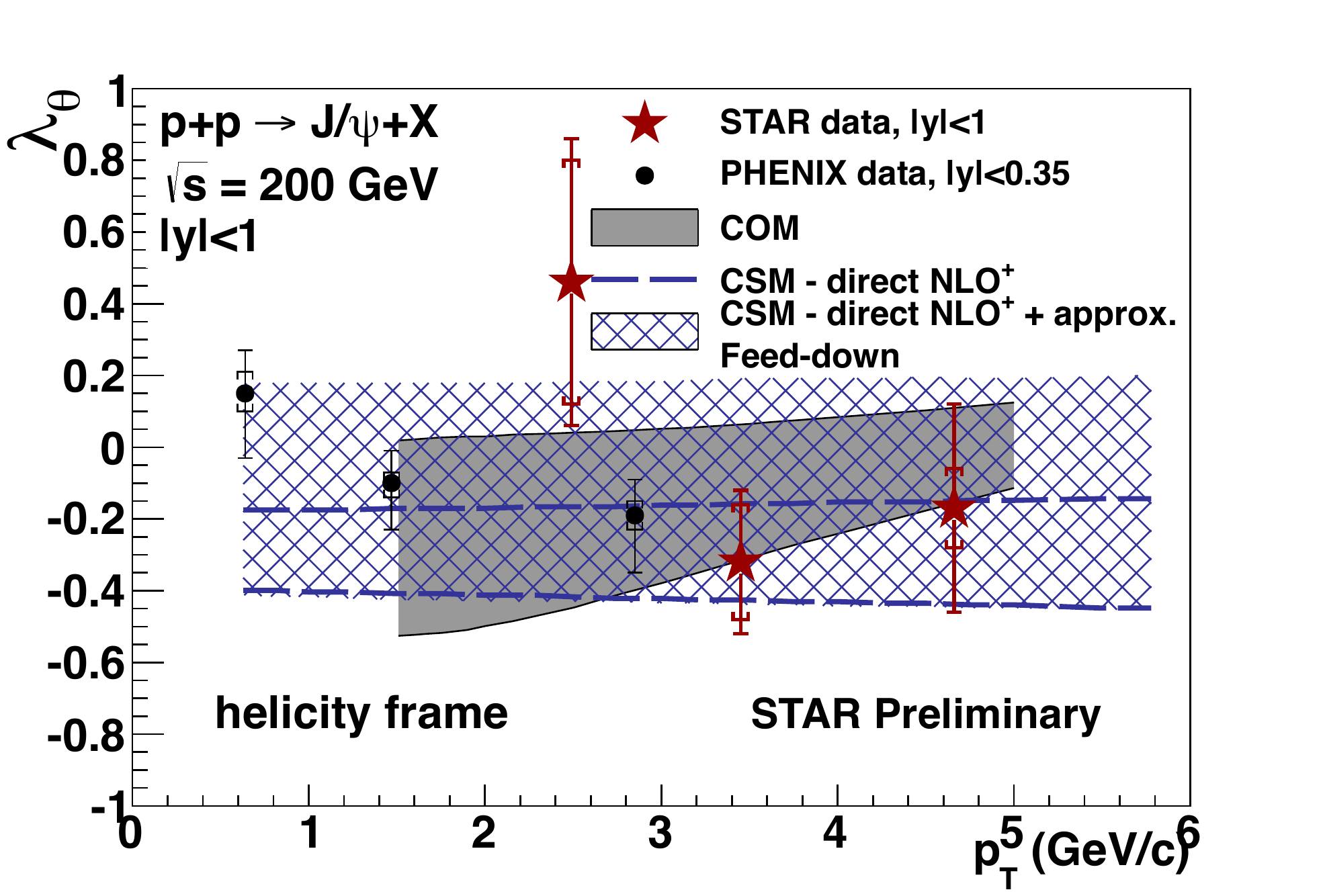}
		\vspace{-15pt}
		\caption{\Jpsi polarization parameter $\lambda_{\theta}$ vs. \pT in $p+p$ collisions. Red stars represent the STAR result.}
		\vspace{-10pt}
		\label{fig:JpsiPolarization}
	\end{minipage}
\end{figure}

Figure \ref{fig:JpsiPtSpectrum_pp} shows \Jpsi transverse momentum spectrum in $p+p$  collisions from year 2009. The new STAR result covers a broad \pT region (0 $<p_{T}<$ 14 GeV/$c$).  The plot also shows predictions from various \Jpsi production models. The Color Evaporation Model (CEM) \cite{CEMJpsi} for prompt \Jpsi can describe the \pT spectrum reasonably well. NLO Non-Relativistic QCD (NRQCD) calculations with color-singlet (CS) and color-octet (CO) transitions \cite{CSCOJpsi} for prompt \Jpsi match the data for $p_{T} > 4$ GeV/c. NNLO* CS model \cite{CSJpsi} for direct \Jpsi production under-predicts the data, but the prediction does not include contributions from feed-down from higher charmonium states and $B$-hadron decays.

Different models of \Jpsi production are able to describe the measured \Jpsi production cross section reasonably well. \Jpsi polarization measurement can help to distinguish among the models since they predict different dependence on \pT for the \Jpsi polarization. \Jpsi polarization in STAR is analyzed in the helicity frame at $|y|<$ 1 and 2 $<p_{T}<$ $\sim$5  GeV/c \cite{STARpolarization}. The \Jpsi polarization parameter $\lambda_{\theta}$ is extracted in three \pT bins and the result is shown in Fig. \ref{fig:JpsiPolarization}. Within current experimental and theoretical uncertainties the obtained transverse momentum dependent $\lambda_{\theta}$ is consistent with the predictions from $NLO^{+}$ Color Singlet Model (CSM) \cite{CSMPolarization} and NRQCD calculations with color octet contributions (COM) \cite{COMPolarization}, and with no polarization.

\vspace{-10pt}
\section{\Jpsi production in $d$+Au and Au+Au and \Jpsi $v_{2}$ in Au+Au collisions at 200 GeV}

\begin{figure}[htbp]
	\begin{minipage}[b]{0.49\linewidth}
	\vspace{-10pt}
		\centering
		\includegraphics[width=0.95\linewidth]{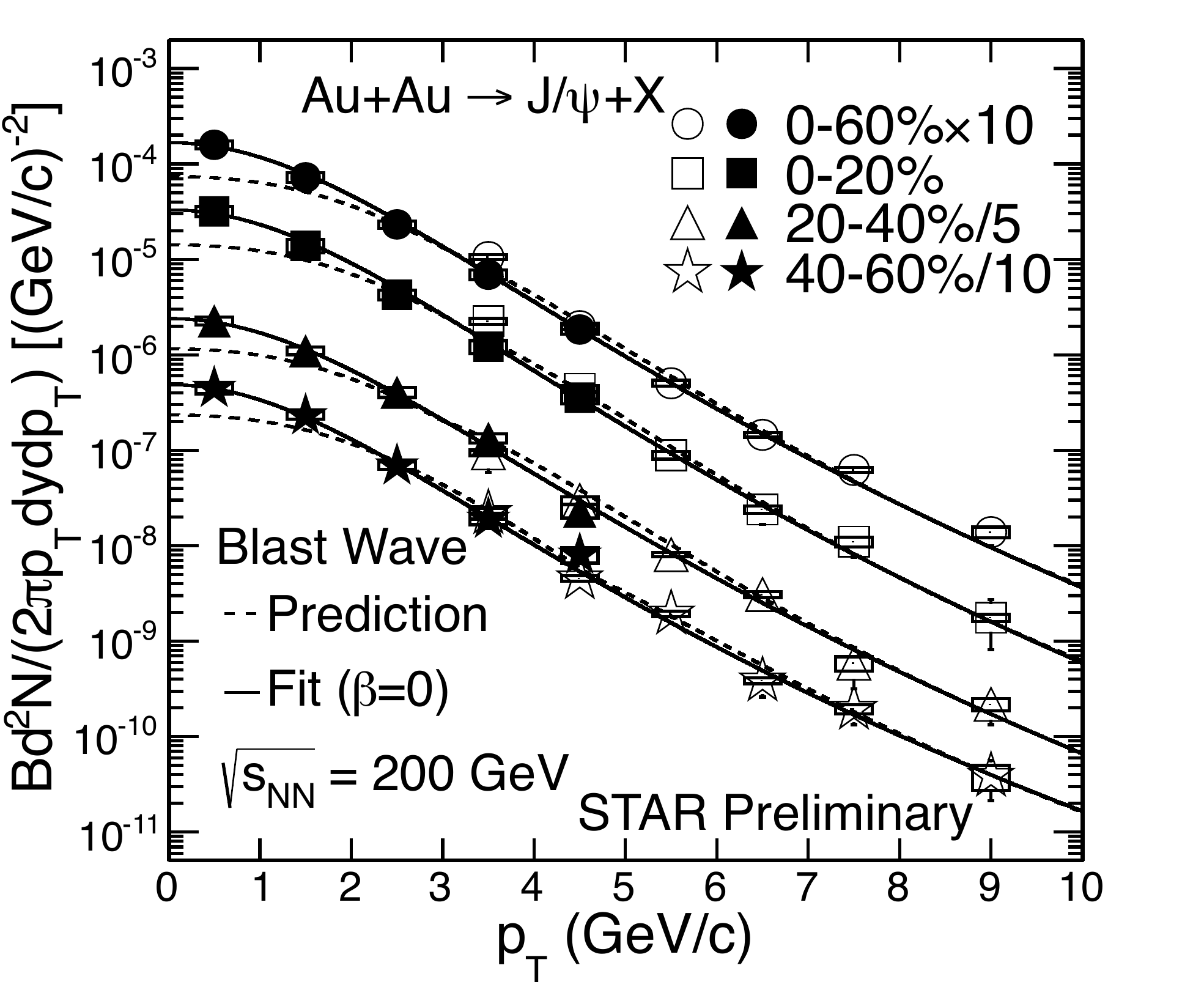}
		\vspace{-15pt}
		\caption{$J/\psi$ \pT spectrum in Au+Au collisions. Full and open symbols represent low-\pT  and  high-\pT \cite{STARhighPtJpsi} measurements, respectively.}
		\vspace{-10pt}
		\label{fig:JpsiPtSpectrum_AuAu}
	\end{minipage}
	\hspace{0.cm}
	\begin{minipage}[b]{0.49\linewidth}
	\vspace{-10pt}
		\centering
		\includegraphics[width=1.\linewidth]{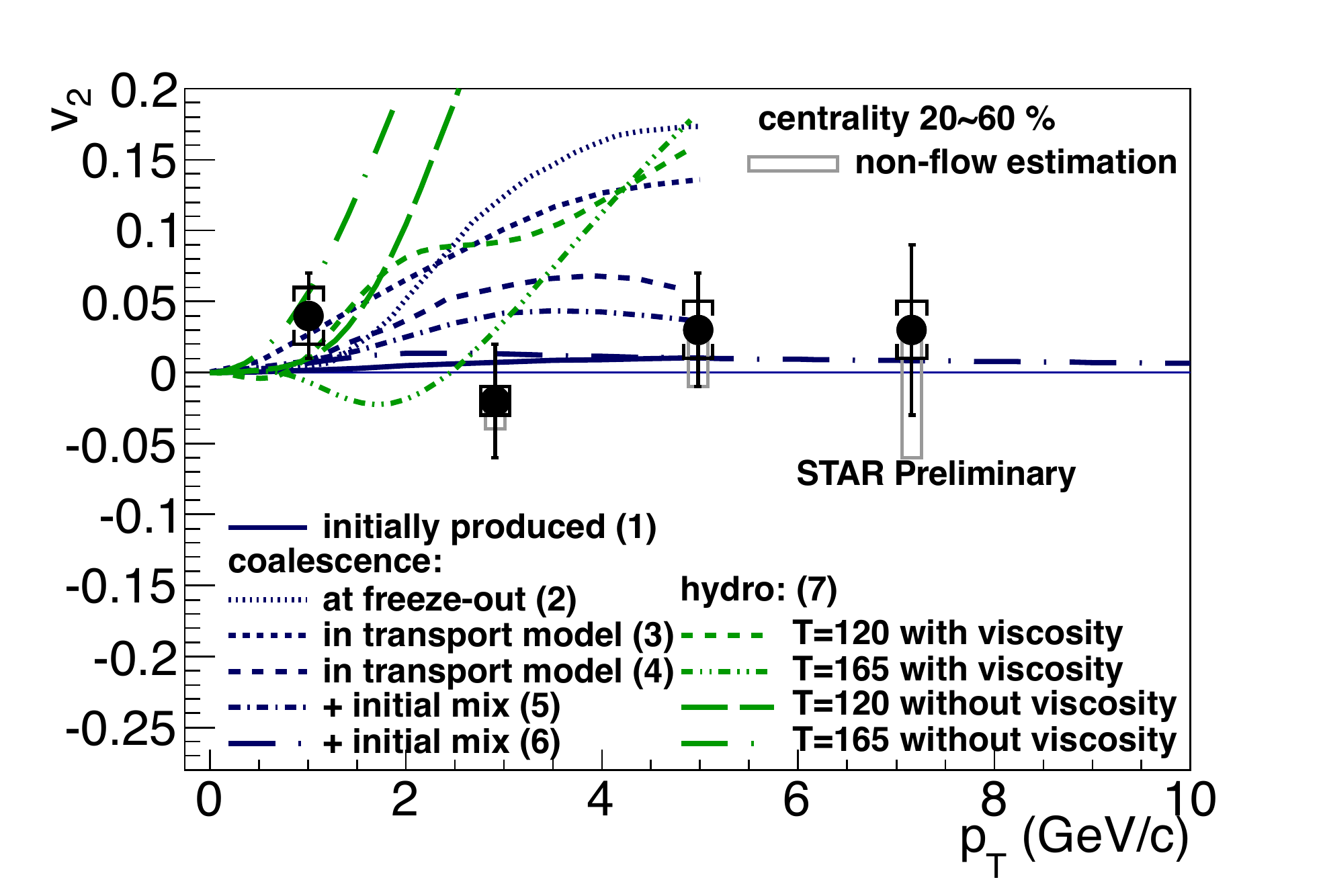}
		\vspace{-15pt}
		\caption{\Jpsi $v_{2}$ vs. \pT for semi-central (20-60\%) Au+Au collisions with different model predictions \cite{v2Model14}-\cite{v2Hydro}.}
		\vspace{-10pt}
		\label{fig:Jpsiv2}
	\end{minipage}
\end{figure}

\Jpsi \pT spectra in Au+Au collisions for different centralities for both low and high \pT (0 $<p_{T}<$ 10 GeV/$c$) are shown in Fig. \ref{fig:JpsiPtSpectrum_AuAu}. The obtained spectra are softer at low \pT than the Tsallis statistics Blast-wave (TBW) model prediction (dashed curves) which assumes that \Jpsi flows like light hadrons \cite{TBW,TBW0}. The data can be described by TBW fit with radial flow velocity $\beta$ fixed to zero (solid curves) \cite{TBW0}. This could be due to a significant contribution from charm quark recombination at low \pT or small \Jpsi radial flow.

\Jpsi $v_{2}$ measurement is a crucial for testing the charm quark recombination effect. Primordial \Jpsi which are produced in the initial hard scattering are expected to carry very little flow, while those that are subsequently created from the recombination of thermalized (anti-)charm quarks will exhibit considerable flow.
\Jpsi $v_{2}$ as a function of \pT for semi-central (20-60\%) Au+Au collisions is shown in Fig. \ref{fig:Jpsiv2} with different model predictions \cite{v2Model14}-\cite{v2Hydro}. The STAR $v_{2}$ result is consistent with zero within the errors. It disfavors the case that \Jpsi is produced dominantly by coalescence from thermalized charm quarks for $p_{T} >$ 2 GeV/$c$ as predicted by, e.g. model (2). Models that assume only initial production of \Jpsi or include both initial production and coalescence process describe the data well.

\begin{figure}[htbp]
	\begin{minipage}[b]{0.49\linewidth}
	\vspace{-10pt}
		\centering
		\includegraphics[width=0.95\linewidth]{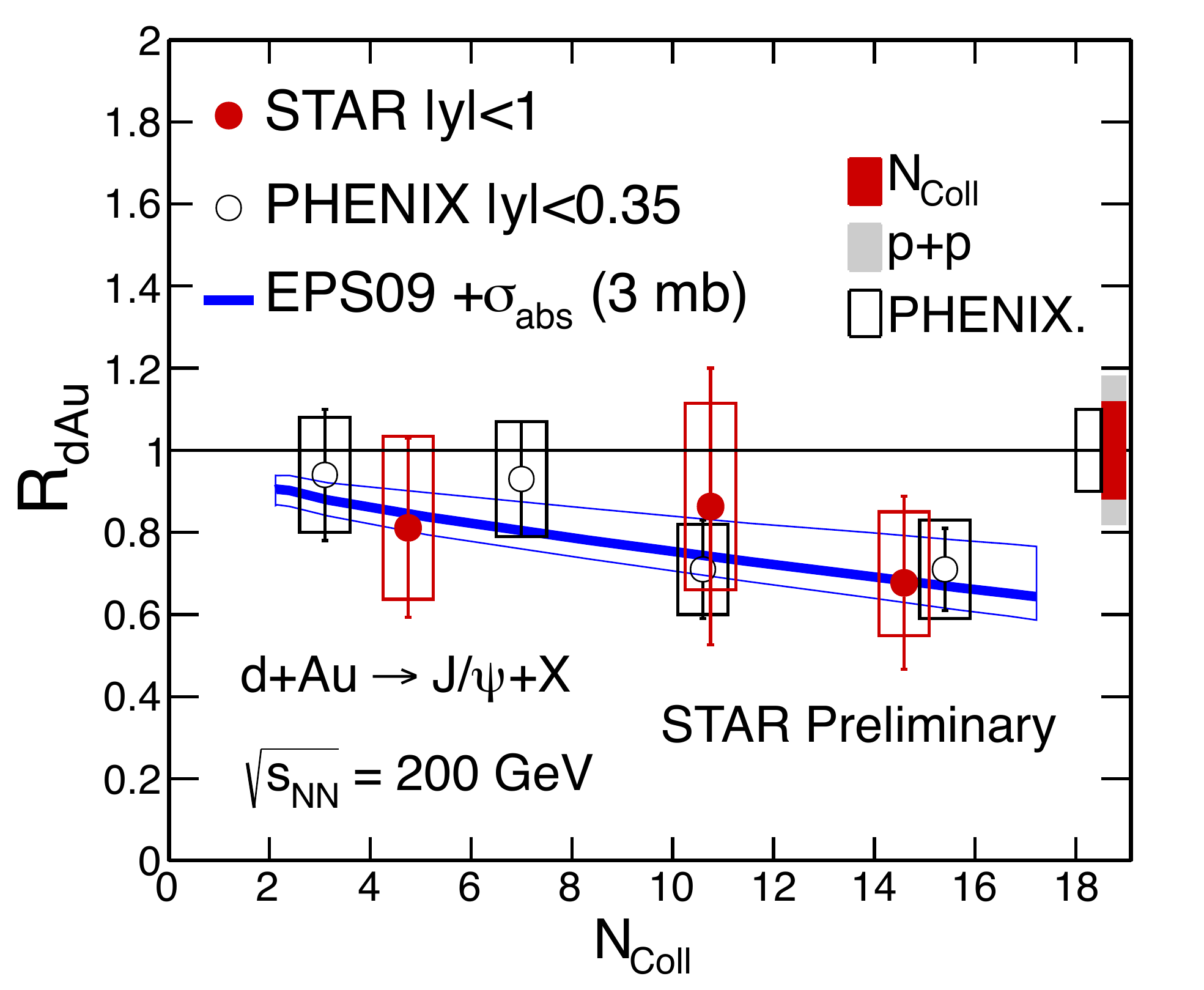}
		\vspace{-15pt}
		\caption{\Jpsi$R_{AA}$ vs. $N_{coll}$ in $d$+Au collisions. Filled circles represent the STAR result.}
		\vspace{-10pt}
		\label{fig:JpsiRaa_dAuNcoll}
	\end{minipage}
	\hspace{0.cm}
	\begin{minipage}[b]{0.49\linewidth}
	\vspace{-10pt}
		\centering
		\includegraphics[width=0.9\linewidth]{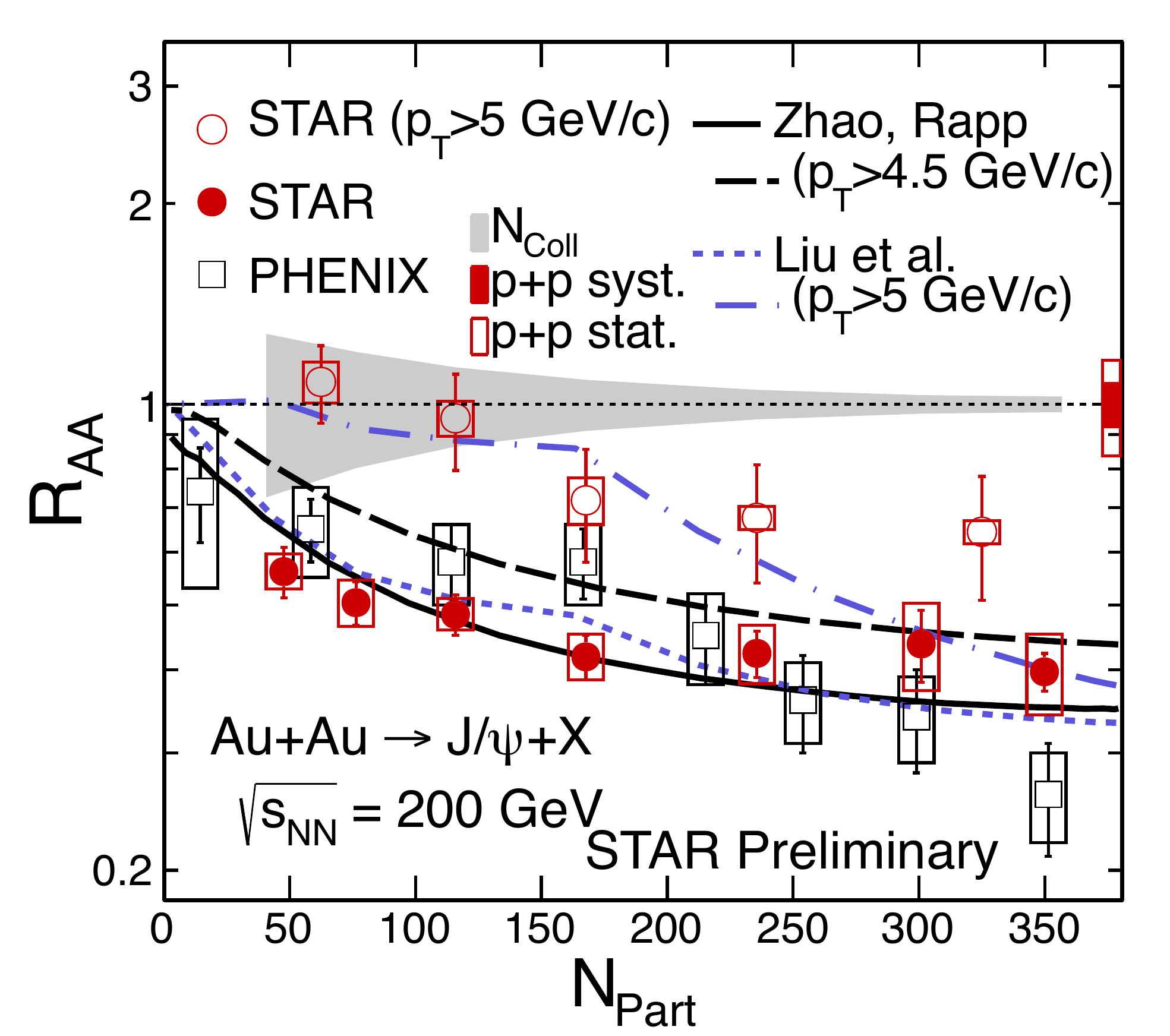}
		\vspace{-15pt}
		\caption{\Jpsi$R_{AA}$ vs. $N_{part}$ in Au+Au collisions. Full and open circles represent low-\pT  and  high-\pT \cite{STARhighPtJpsi} measurements, respectively.}
		\vspace{-10pt}
		\label{fig:JpsiRaa_AuAuNpart}
	\end{minipage}
\end{figure}

Figure \ref{fig:JpsiRaa_dAuNcoll} shows \Jpsi $R_{AA}$ in $d$+Au collisions as a function of $N_{coll}$ for $p_{T} <$ 5 GeV/c. The data are in good agreement with a model prediction using EPS09 parametrization of nuclear parton distribution functions for the shadowing and a \Jpsi nuclear absorption of $\sigma_{abs}$ = 3 mb \cite{EPS09}. The $\sigma_{abs}$ = 2.8$^{+3.5}_{-2.6}(stat.) ^{+4.0}_{-2.8}(syst.) ^{+1.8}_{-1.1}(EPS09)$ mb was obtained from a fit to the data.

\Jpsi $R_{AA}$ in Au+Au collisions as a function of $N_{Part}$ at low and high \pT \cite{STARhighPtJpsi} is shown in Fig. \ref{fig:JpsiRaa_AuAuNpart}. The observed \Jpsi suppression increases with a collision centrality and decreases towards higher \pT across the centrality range. At high \pT we observe suppression only for central collisions.  The results are compared to two models that include primordial \Jpsi production (with the color screening effect and CNM effects) and the regeneration from charm quarks \cite{JpsiAuAuLiu,JpsiAuAuZhao}. Low-\pT data agrees with both model predictions. At high \pT Liu et al. model \cite{JpsiAuAuLiu} describes the data reasonably well while Zhao and Rapp model \cite{JpsiAuAuZhao} underpredicts the $R_{AA}$ for $N_{Part}>$ 70.

\vspace{-10pt}
\section{$\Upsilon$ production in $p+p$ and Au+Au collisions at 200 GeV}

\begin{figure}[htbp]
	\begin{minipage}[b]{0.49\linewidth}
	\vspace{-10pt}
		\centering
		\includegraphics[width=1.\linewidth]{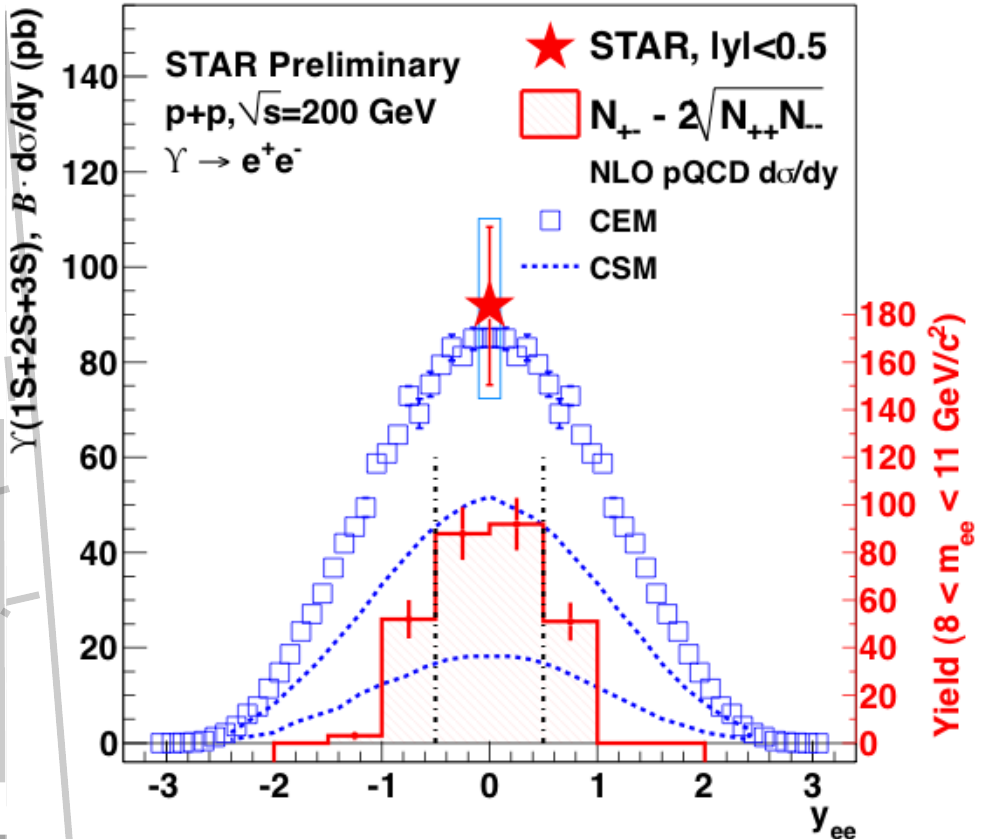}
		\vspace{-15pt}
		\caption{$\Upsilon$(1S+2S+3S) cross section vs. rapidity in $p+p$ collisions.}
		\vspace{-10pt}
		\label{fig:UpsilonCrossSection_ppY}
	\end{minipage}
	\hspace{0.cm}
	\begin{minipage}[b]{0.49\linewidth}
	\vspace{-10pt}
		\centering
		\includegraphics[width=1.\linewidth]{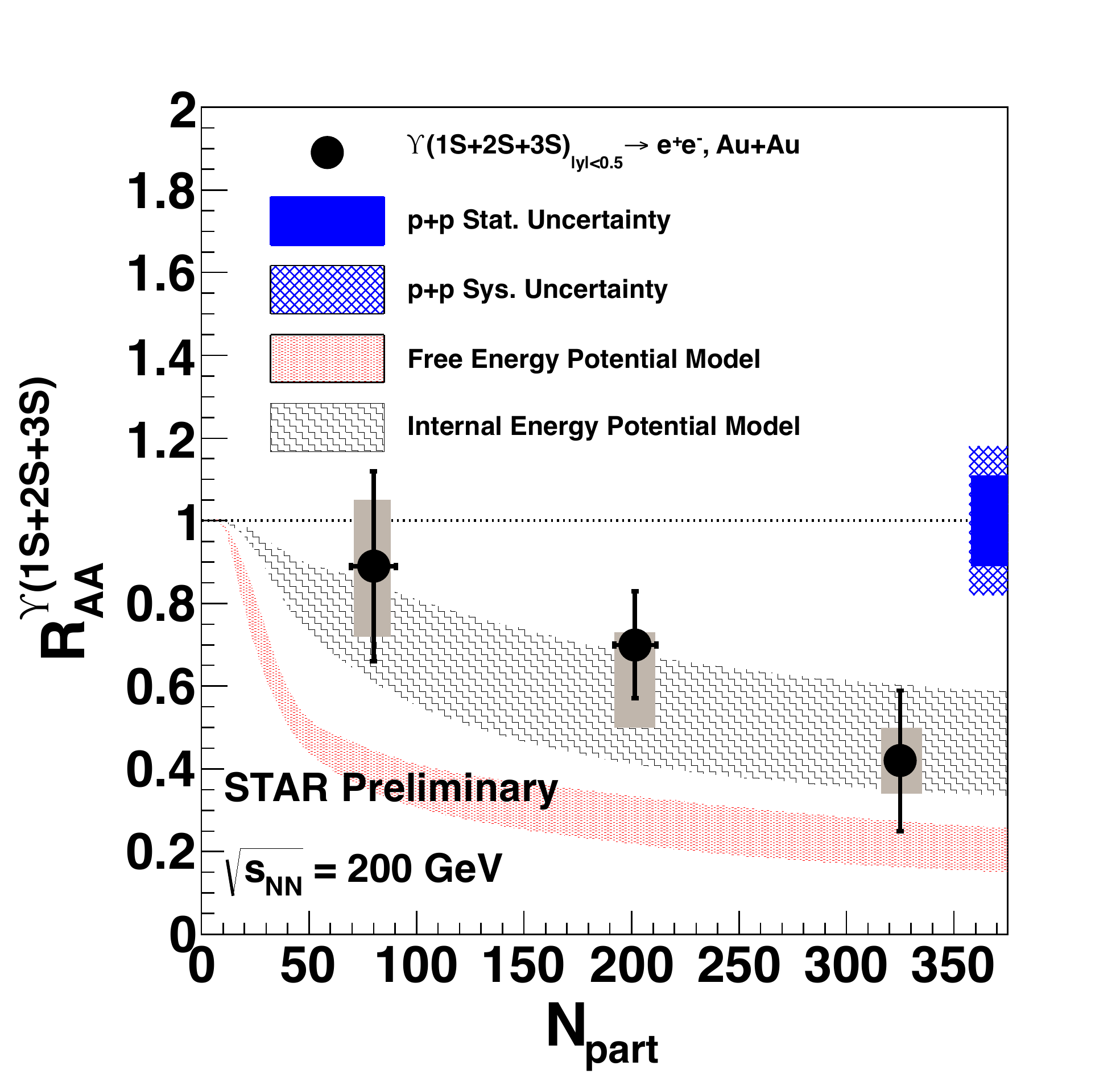}
		\vspace{-15pt}
		\caption{$\Upsilon$(1S+2S+3S) $R_{AA}$ vs. $N_{part}$ in Au+Au collisions.}
		\vspace{-10pt}
		\label{fig:UpsilonRaa_AuAuNpart}
	\end{minipage}
\end{figure}

STAR has improved precision of the $p+p$ $\Upsilon$(1S+2S+3S)$\rightarrow e^{+}e^{-}$ cross section measurement using the 2009 data with enhanced statistics. Figure \ref{fig:UpsilonCrossSection_ppY} shows the new STAR result as a function of rapidity with two model predictions, CEM \cite{UpsilonCEM} and CSM \cite{UpsilonCSM}. 
The obtained cross section is consistent with the NLO  pQCD calculations.

Figure \ref{fig:UpsilonRaa_AuAuNpart} shows $\Upsilon$(1S+2S+3S) $R_{AA}$ as a function of $N_{part}$ in Au+Au collisions at mid-rapidity. Results are compared with two predictions of dynamic model with fireball expansion and quarkonium feed-down \cite{Strickland}. The calculations include variation of initial $\eta/S$ and $T_{0}$. The observed suppression at central collisions is consistent with the prediction of complete melting of 3S state and very strong suppression of 2S state from the model that uses internal energy as the heavy quark potential.

\vspace{-10pt}
\section{Summary}

In summary, the recent results of STAR \Jpsi and $\Upsilon$ measurements in $p+p$, $d$+Au and Au+Au collisions at  $\sqrt{s_{NN}}$ = 200 GeV are shown.  \Jpsi $R_{dAu}$ agrees with the model using EPS09 + $\sigma_{abs}^{J/\psi}$ (3 mb). \Jpsi $R_{AuAu}$ decreases with centrality and increases with $p_{T}$. At high \pT suppression is only seen is central collisions. \Jpsi $v_{2}$ was found to be consistent with zero, it disfavors the case that \Jpsi is produced dominantly by coalescence of thermalized charm quarks at $p_{T} >$ 2 GeV/c. $\Upsilon$(1S+2S+3S) Au+Au results are consistent with the model that predicts complete melting of 3S and a strong 2S suppression.


\vspace{-10pt}
\section*{References}

\begin{thebibliography}{00} 

\bibitem{MatsuiSatz} T.Matsui, H.Satz, Phys. Lett. B 178, 416 (1986)
\bibitem{STARhighPtJpsi} L.Adamczyk et al., arXiv:1208.2736v1
\bibitem{CEMJpsi} A.D.Frawley, T.Ullrich, R. Vogt, Phys. Rept. 462, 125 (2008), and R.Vogt private communication (2009)
\bibitem{CSCOJpsi} Y.-Q.Ma, K.Wang, K.T.Chao, Phys. Rev. D84, 51 114001 (2011), and private communication (2012)
\bibitem{CSJpsi} P.Artoisenet et al., Phys. Rev. Lett. 101, 152001 (2008) and J.P.Lansberg private communication (2009)
\bibitem{TBW} Z.Tang eta al., arXiv:1101.1912 (2011)
\bibitem{TBW0} Z.Tang et al., Phys. Rev. C79, 051901 (2009)
\bibitem{STARpolarization} B.Trzeciak (for the STAR Collaboration), Acta Physica Polonica B Proceedings Supplement Vol.5 page 549 (2012)
\bibitem{CSMPolarization} J.P.Lansberg, Phys. Lett. B 695, 149-156 (2011)
\bibitem{COMPolarization} H.S.Chung, C.Yu, S.Kim, J.Lee, Phys. Rev. D 81, 014020 (2010)
\bibitem{PHENIXPolarization} A.Adare et al., Phys. Rev. D 82, 012001 (2010)
\bibitem{EPS09} K.Eskola, H.Paukkunena, P.Ruuskanen, Nucl. Phys. A830, 599 (2009), R.Vogt, Phys. Rev. C81, 044903 (2010)
\bibitem{JpsiAuAuLiu} Y.Liu, Z.Qu, N.Xu, P.Zhuang, Phus. Lett. B678, 72 (2009)
\bibitem{JpsiAuAuZhao} X.Zhao, R.Rapp, Phys. Rev. C82, 064905 (2010)
\bibitem{v2Model14} L.Yan,P.Zhuang, N.Xu, Phys. Rev. Lett. 97, 232301 (2006) - \Jpsi $v_{2}$ Models (1) (4)
\bibitem{v2Model2} V.Greco, C.M.Ko, R.Rapp, Phys. Lett. B595, 202 (2004) - \Jpsi $v_{2}$ Models (2)
\bibitem{v2Model3} L.Ravagli, R.Rapp, Phys. Lett. B655, 126 (2007) - \Jpsi $v_{2}$ Models (3)
\bibitem{v2Model5} X.Zhao, R.Rapp, 24th Winter Workshop on Nuclear Dynamics - \Jpsi $v_{2}$ Models (5)
\bibitem{v2Model6} Y.Liu, N.Xu, P.Zhuang, Nucl. Phys. A834, 317 (2010) - \Jpsi $v_{2}$ Models (6)
\bibitem{v2Hydro} U.W.Heinz, C.Shen, (2011), private communication - \Jpsi $v_{2}$ Models (7)
\bibitem{UpsilonCEM} R.Vogt, Phys. Rep. 462125 (2008)
\bibitem{UpsilonCSM} J.P.Lansberg, S.Brodsky, PRD 81, 051502 (2010)
\bibitem{Strickland} M.Strickland, D. Bazow, arXiv:1112.2761v4 (2012)

\end{thebibliography}
\end{document}